\documentclass[10pt,conference]{IEEEtran}
\usepackage{cite}

\ifCLASSINFOpdf 
\usepackage[pdftex]{graphicx}
\else
\usepackage[dvips]{graphicx}
\fi

\usepackage[cmex10]{amsmath}
\usepackage{algorithmic}
\usepackage[ruled,norelsize]{algorithm2e}

\usepackage{array}

\usepackage{mdwmath}
\usepackage{mdwtab}
\usepackage{todonotes}

\usepackage{eqparbox}

\ifCLASSOPTIONcompsoc
\usepackage[caption=false,font=normalsize,labelfont=sf,textfont=sf]{subfig}
\else
\usepackage[caption=false,font=footnotesize]{subfig}
\fi

\usepackage{fixltx2e}
\usepackage{url}
\usepackage{amsfonts}
\usepackage{amssymb}
\usepackage{amsthm}
\usepackage{bm}
\usepackage[ruled]{algorithm2e}
\usepackage[utf8]{inputenc}
\usepackage{booktabs}
\usepackage{url}
\usepackage{cite}
\usepackage{graphicx}
\usepackage{tabularx}
\usepackage[caption=false,font=footnotesize]{subfig}

\usepackage{epstopdf}

\newcounter{tempEquationCounter} 
\newcounter{thisEquationNumber}

\begin{document}
%


\title{Powering Net-Zero 6G: Packetized Energy Management for Grid-Interactive Telecom Infrastructure}
\vspace{-10mm}

%

\author{
\IEEEauthorblockN{Adnan Aijaz, Xinyi Lin}
\IEEEauthorblockA{
\text{Bristol Research and Innovation Laboratory, Toshiba Europe Ltd., Bristol, United Kingdom}\\
}
\vspace{-10mm}
}   

\markboth{IEEE XYZ Magazine -- Submitted for Publication}%
{Shell \MakeLowercase{\textit{et al.}}: Bare Demo of IEEEtran.cls for Journals}
%


\maketitle
\begin{abstract}
\boldmath
The transition to net-zero 6G requires energy-management approaches that go beyond conventional RAN efficiency mechanisms. As future networks integrate AI-native operation, edge intelligence, dense deployments, renewables, and storage, the RAN will become both a growing power consumer and a source of distributed energy flexibility. This paper introduces \emph{packetized energy management} (PEM) as a framework for transforming 6G infrastructure into energy-aware, grid-interactive assets. PEM represents flexible demand as schedulable energy packets that can be admitted, deferred, or reshaped according to local constraints, renewable availability, carbon intensity, price, and communication priorities. We present a PEM-enabled base-station model, a RAN architecture for PEM integration, and the telecoms virtual power plant (VPP) concept for aggregating PEM-enabled sites. Simulation results demonstrate PEM’s potential for peak-aware operation, improved renewable utilization, and outage-resilient service continuity. The paper also discusses open challenges for telco-energy co-design.
\end{abstract}

%
\IEEEpeerreviewmaketitle

\section{Introduction}
\IEEEPARstart{S}{ustainability} is taking the center stage in various sectors including telecoms. Achieving sustainability in telecoms necessitates a multifaceted approach which revolves around energy efficiency, green network design, integration of renewable energy sources, equipment circularity, and green supply chain management. 
Embracing net-zero in current/future wireless systems is a crucial step toward a sustainable telecoms ecosystem. There is a growing focus on carbon footprint reduction of mobile/cellular network, from energy-efficient 5G operation to net-zero 6G design, as operators worldwide commit to net-zero targets by 2050 or earlier \cite{GSMA_netzero}. 

On the path to net-zero in telecoms, the low-hanging fruit is to configure the radio access network (RAN) with energy saving features in time, frequency, or spatial domains \cite{RAN_EE}, potentially improving energy efficiency by 10-12\% \cite{NGMN_EE2}. However, energy efficiency alone is insufficient; realizing net-zero necessitates cross-sector innovation beyond the telecoms industry; a synergistic integration of energy and telecoms sectors can play a pivotal role in realizing the vision of net-zero by design 6G. The transition toward AI-native 6G, edge intelligence, and accelerated computing within or near the RAN is expected to increase the power consumption of base stations and edge nodes. This reinforces the need for innovative energy-management approaches that can dynamically reduce, shift, and optimize site-level energy usage beyond energy efficiency.  Establishing dynamic, bidirectional energy flows between telecom networks and the wider power grid can unlock mutual value and support measurable carbon-emission reduction \cite{GSMAIntelligence_VPP_2024}.


Telecom infrastructure is increasingly viewed as a distributed flexibility resource. Off-grid and weak-grid RAN sites can use renewables and storage to reduce diesel dependence, while grid-connected sites can repurpose backup batteries as controllable storage. This shift is already emerging: Elisa, a Finnish telecommunications operator, has launched a distributed energy storage initiative targeting 150~MWh of mobile base station backup capacity. Large-scale studies further show that coordinated network energy management and solar integration can support decarbonization, with one 5G study in China estimating that such measures could deliver more than 50\% of the sector's net-zero goal \cite{li2023carbon}. Coordinated RAN energy assets can therefore improve resilience, reduce outage-related losses, and provide flexibility to the power system.


This paper advocates \emph{packetized energy management} (PEM) for next-generation mobile/cellular networks. PEM models flexible demand as schedulable energy packets that can be admitted, deferred, or reshaped based on grid conditions, renewable availability, priorities, and local constraints. This fits 6G RAN infrastructure, where cooling, rectifiers, batteries, auxiliary loads, and delay-tolerant edge workloads can be coordinated without disrupting mission-critical functions. PEM thus enables energy-aware, grid-interactive RAN operation and supports telecoms virtual power plant (VPP) aggregation for energy, cost, carbon, and grid-service optimization.



Prior work on telecom sustainability has mainly addressed RAN energy efficiency, renewable integration, and carbon reduction, while PEM has largely been studied for power systems and distributed energy resources. Its role as a telecom-native energy-control framework remains unexplored, especially for service continuity, QoS preservation, outage resilience, and grid-facing aggregation. To the best of our knowledge, this is among the first works to investigate PEM for mobile/cellular networks and position PEM-enabled 6G RAN as a building block for telecoms VPP.
The key contributions are as follows:
\begin{itemize}
    \item We introduce a PEM-enabled 6G RAN framework that identifies packetizable base station loads, local PEM control functions, and energy-aware operating modes.
    \item We develop an architectural view of PEM integration with 6G RAN management, including telemetry models, APIs, interfaces, and orchestration functions.
    \item We propose the Telecoms VPP concept, showing how PEM-enabled base stations can be aggregated as grid-interactive energy resources.
    \item We evaluate PEM through fleet-level and outage-resilience studies, quantifying its impact on energy, carbon, cost, renewable utilization, peak control, and UE service continuity.
    \item We discuss open challenges in QoS-aware packetization, interoperability, standardization, market participation, and scalable telco-energy co-design.
\end{itemize}


We first review PEM fundamentals and its relevance to net-zero 6G and grid-interactive telecom operation. We then present the PEM-enabled RAN architecture and Telecoms VPP concept, followed by fleet-level and outage-resilience evaluations, key challenges, and concluding remarks.

\section{What is Packetized Energy Management?}
PEM \cite{PEM_intro1} is a demand-side load control technique that models energy as discrete, time-bound, and power-constrained packets requested by devices from a controller. Inspired by the concepts of packetization and randomization, which underpin scalable and decentralized operation of the Internet, PEM empowers the power/electric grid for real-time, scalable coordination of distributed energy resources (DERs) by allocating energy in small, fixed-duration and fixed-power packets. It also improves the reliability of the power grid  while maximizing the use of renewable energy and without the need for scaling up grid infrastructure. 
The basic operation of PEM can be summarized in the following key steps.  

\begin{itemize}
\item \emph{Local state assessment} -- A DER measures its local energy state, e.g., its state-of-charge (SoC). 

\item \emph{Packetized mode} -- If the measured SoC is within a predefined range, the DER operates in the packetized mode, making probabilistic requests to either consume or inject energy at a fixed rate for a specified duration—thereby generating an energy packet. When the SoC is outside this range, the DER exits packetized mode and reverts to its default control mode, remaining in that mode until the SoC returns within the predefined limits.

\item \emph{Request Evaluation} -- The aggregator or the VPP evaluates each energy packet request from the DER based on prevailing grid/system conditions and market dynamics, and either accepts or rejects it. If the request is rejected, the DER immediately returns to the state assessment phase. If accepted, the DER proceeds to consume or inject energy for the defined epoch, after which it resumes state assessment.

\end{itemize}

Further details on PEM are available in \cite{PEM_intro1, PEM_intro2}.

\section{Why PEM for 6G?}

PEM supports 6G decarbonization by converting site-level energy flexibility into a controllable resource. Unlike conventional RAN energy-saving methods, which mainly optimize communication functions, PEM coordinates flexible energy demand across telecom sites according to renewable availability, grid conditions, carbon intensity, and service priorities. Its relevance to 6G is summarized as follows.

\begin{itemize}

\item \textbf{Grid-Interactive Infrastructure} -- PEM enables non-critical loads such as HVAC, rectifiers, batteries, auxiliary systems, and delay-tolerant workloads to request schedulable energy packets. These packets can be admitted, deferred, or reshaped to reduce demand during high-carbon, high-price, or grid-constrained periods without interrupting mission-critical RAN functions.

\item \textbf{Renewable-Aware Operation} -- Net-zero telecom operation depends on using on-site renewables effectively. PEM aligns flexible demand with solar/wind availability by prioritizing battery charging, pre-cooling, and deferrable workloads during renewable surplus, while delaying non-critical consumption when output declines. This improves self-consumption, reduces curtailment, and limits reliance on fossil-based grid power or diesel backup.

\item \textbf{Energy-Aware Network Management} -- PEM introduces an energy-control layer that interacts with RAN orchestration and management functions. By exposing site energy states, flexibility envelopes, and power budgets, it allows energy cost, carbon intensity, and resilience to be considered alongside capacity, coverage, and latency, making energy an active control variable in 6G operation.

\item \textbf{Device-Level Load Shaping} -- PEM can extend to UEs, IoT devices, and edge terminals whose charging, sensing, computation, or data transfer tasks have timing flexibility. Aggregated across large device populations, these micro-loads can support demand shaping and renewable alignment, extending sustainability beyond infrastructure.

\item \textbf{RAN as an Energy Asset} -- PEM helps transform the RAN from a passive load into a distributed energy asset. During grid instability or outages, it can prioritize essential communication loads, defer non-critical packets, and allocate stored or renewable energy to preserve connectivity\footnote{Power disruptions can cost telecom operators an estimated USD~3k--5k per site per hour in lost revenue and service degradation \cite{mckinsey2021}.}. This enables graceful degradation, reduces downtime, and strengthens service continuity.

\item \textbf{Data-Driven Orchestration} -- 
PEM generates fine-grained telemetry on energy packets, load states, battery SOC, renewable output, and flexibility availability. These data support predictive scheduling, carbon accounting, and AI-assisted orchestration, while open interfaces and standardized data models enable integration with RAN platforms, energy markets, and grid operators.

\end{itemize}

\section{What PEM-Enabled 6G Brings?}

From the power-grid perspective, PEM-enabled 6G creates a new class of distributed, communication-aware energy resources. Base stations, edge nodes, batteries, renewables, and connected devices can collectively provide controllable demand and distributed storage beyond the scale of conventional commercial loads. The key value for the energy sector is summarized as follows.

\begin{itemize}

\item \textbf{Aggregated Distributed Flexibility} -- PEM-enabled telecom sites can pool flexible demand and storage capacity. Packetized control coordinates many small site-level actions into aggregate flexibility for demand response, peak shaving, and renewable balancing.

\item \textbf{Grid Stability and Resilience} -- By adjusting the timing and magnitude of telecom energy use, PEM-enabled networks can relieve local grid stress and support stability under renewable variability. Their spatial distribution and fast response make telecom sites suitable for decentralized grid-support functions, subject to service constraints.

\item \textbf{Improved Renewable Utilization} -- PEM aligns telecom demand with renewable generation at site and grid levels. During renewable surplus, sites can absorb energy through storage charging or flexible loads; during constrained supply, non-critical packets can be deferred. This reduces curtailment and supports higher variable-renewable penetration.

\item \textbf{Grid Visibility and Predictive Intelligence} -- PEM telemetry provides high-resolution visibility into load flexibility, battery state, thermal limits, and packet acceptance across telecom sites. These data can support predictive dispatch, anomaly detection, and more reliable flexibility forecasting.

\item \textbf{Market Participation and Telecoms VPPs} -- Aggregated PEM-enabled sites can form a Telecoms VPP, enabling operators to monetize flexibility while supporting grid services. Market participation requires models that capture telecom-specific constraints, including coverage obligations, resilience reserves, and service-level agreements.

\item \textbf{Sector Coupling and Co-Evolution} -- PEM enables deeper coupling between telecom and energy systems. The grid gains a responsive digital flexibility layer, while telecom networks benefit from carbon-aware energy sourcing, lower operating costs, and improved resilience, supporting grid-interactive net-zero 6G infrastructure.

\end{itemize}

\section{PEM for 6G RAN Infrastructure}

\subsection{PEM-Enabled Base Station}

\begin{figure}
\centering
\includegraphics[width=\columnwidth]{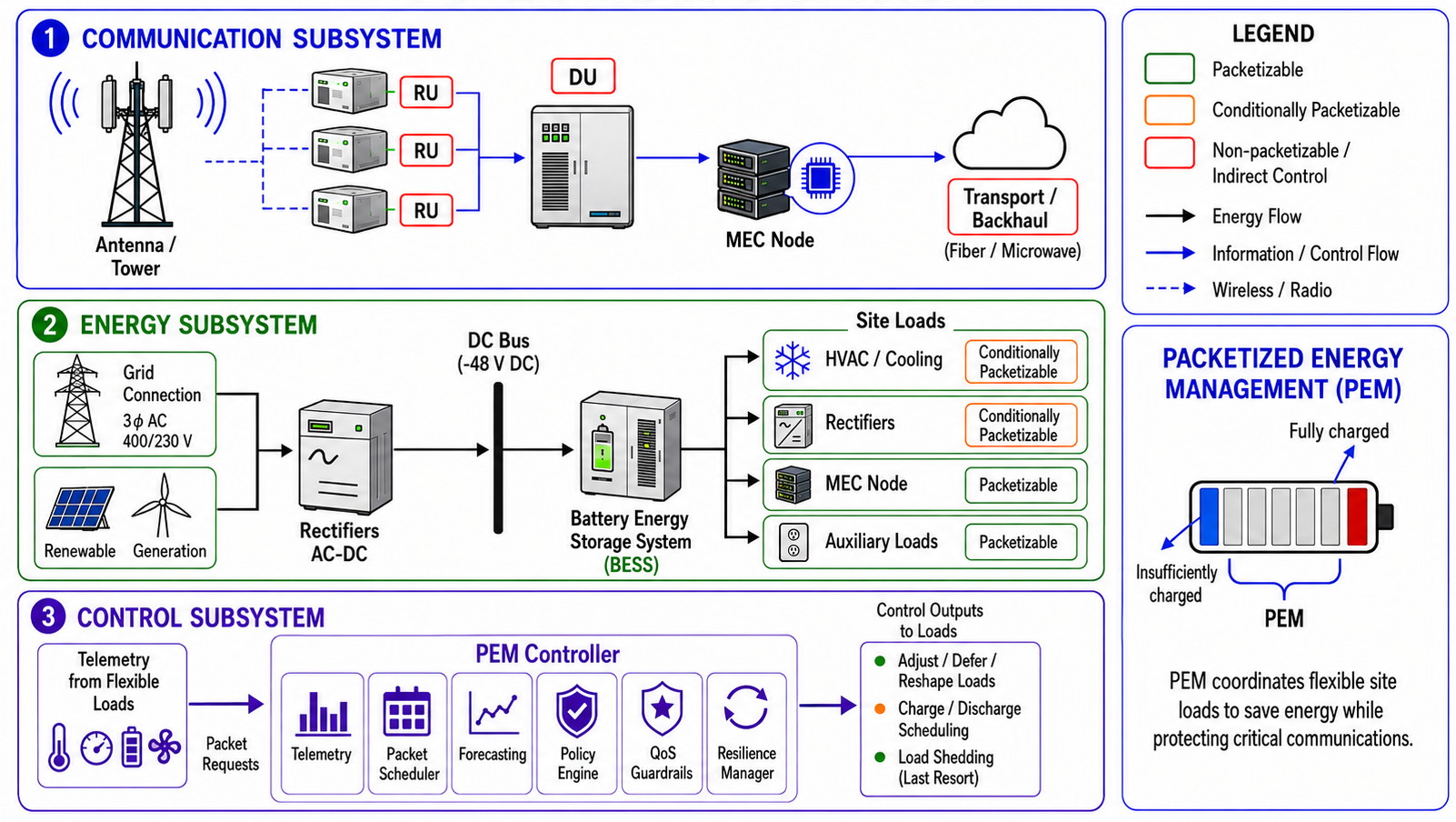}
\caption{PEM-enabled 6G base station. }
\label{PEM_BS}
\vspace{-1.5em}
\end{figure}

A PEM-enabled base station augments conventional 6G RAN infrastructure with local energy intelligence for coordinating power consumption, on-site generation, and storage. As shown in Fig.~\ref{PEM_BS}, the architecture comprises three subsystems: the communication subsystem, including RUs, DUs, and transport; the energy subsystem, including rectifiers, BESS, renewables, and thermal management; and the control subsystem, centered on the PEM controller.

The PEM controller virtualizes the site energy state using telemetry such as power, voltage, current, battery SOC/SOH, temperature, airflow, renewable output, and grid signals. From these inputs, it derives feasible operating envelopes subject to DC bus limits, thermal constraints, backup-energy reserves, and communication-service requirements. Flexible subsystems then issue energy-packet requests characterized by power, duration, priority, and timing flexibility. The controller admits, defers, or reshapes these packets based on local constraints, forecasts, carbon intensity, price, renewable availability, and higher-layer Telecoms VPP directives.

Operationally, PEM performs three main functions. First, it shapes flexible loads such as HVAC, rectifiers, auxiliary loads, and non-critical edge workloads. Second, it schedules BESS charging and discharging to support renewable utilization, peak control, and resilience. Third, it enforces safety and QoS guardrails so that mission-critical communication functions remain continuously powered.

A key design principle is the separation of non-packetizable and flexible loads. RUs, DUs, synchronization, and transport are latency- and reliability-critical and therefore cannot be directly packetized; their power can only be influenced indirectly through RAN mechanisms such as carrier activation, sleep modes, traffic steering, or admission control. In contrast, HVAC, rectifiers, BESS, renewables, auxiliary loads, and delay-tolerant edge workloads provide the flexibility needed for packetized control.

PEM also supports predictive and resilient operation. Short-term forecasts of traffic, renewable generation, thermal behavior, and energy prices allow pre-cooling, proactive battery scheduling, and constraint-aware packet admission. During grid disturbances, the controller can enter an islanded mode, prioritize essential communication loads, shed or defer non-critical packets, and use stored or renewable energy to extend service continuity. Thus, the PEM-enabled base station becomes the local building block for higher-layer aggregation and Telecoms VPP operation.

\begin{table*}[t]
\centering
\caption{Packetization capability of base-station loads}
\label{tab:bs_packetization}
\scriptsize
\setlength{\tabcolsep}{3.5pt}
\renewcommand{\arraystretch}{0.92}
\begin{tabularx}{\textwidth}{p{2.45cm}p{2.65cm}p{1.55cm}X}
\hline
\textbf{Subsystem} &
\textbf{Load or asset} &
\textbf{Packetizable?} &
\textbf{Remarks} \\
\hline
RAN functions &
RU, DU/baseband, transport, synchronization &
No &
Mission-critical and possibly battery-backed, but not directly packetized \\

Traffic-dependent RAN power &
Dynamic RF/baseband load &
Indirect &
Controlled only through RAN actions such as carrier control, sleep modes, traffic steering, or admission control \\

HVAC and thermal management &
Compressors, fans, cabinet cooling &
Conditional &
Duty-cycled, deferred, or pre-cooled within thermal limits, mainly at macro and metro sites \\

Rectifiers and power conversion &
AC/DC modules, DC bus supply &
Conditional &
Modular staging and load sharing support efficiency and peak-aware operation \\

Battery energy storage &
BESS charge/discharge &
Yes &
Main temporal flexibility source and backup supply for critical non-packetizable loads \\

Renewable interface &
PV/wind and inverter output &
Conditional &
Generation is weather-driven, but curtailment, export, local absorption, and charging are controllable \\

Edge compute &
MEC or AI workloads &
Conditional &
Delay-tolerant workloads can be deferred or degraded, mainly at macro and metro edge sites \\

Flexible auxiliary loads &
Lighting, non-critical sensing, auxiliary fans &
Yes &
Low-priority loads suitable for deferral or shedding with minimal service impact \\
\hline
\end{tabularx}
\vspace{-4mm}
\end{table*}

\subsection{RAN Architecture for PEM in 6G}

\begin{figure}
\centering
\includegraphics[width=\columnwidth]{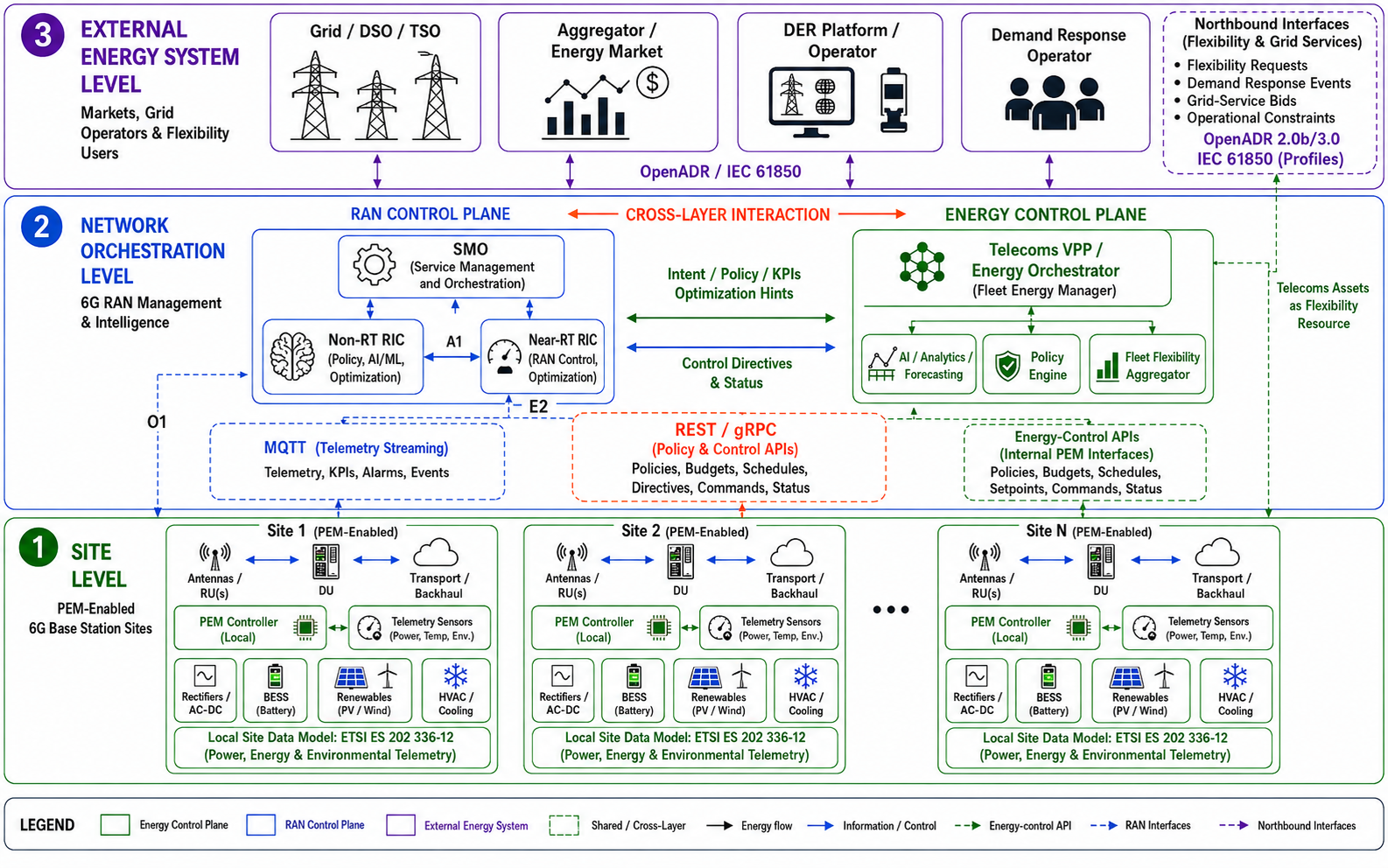}
\caption{RAN architecture to support PEM in 6G.}
\label{PEM_arch}
\vspace{-1.5em}
\end{figure}

Realizing PEM in 6G requires architectural enhancements that embed energy awareness into the RAN control and management framework. As shown in Fig.~\ref{PEM_arch}, this can be achieved through an energy control plane that operates alongside the conventional RAN control plane and connects local PEM controllers with higher-layer orchestration systems and external energy platforms. The objective is to enable cross-layer coordination between communication performance and energy management without compromising latency, reliability, or service continuity.

At the site level, the PEM controller interfaces with the energy subsystem through southbound protocols such as Modbus, SNMP, BACnet, or IEC-based interfaces, depending on vendor equipment and deployment context. Standardized information models, such as ETSI ES 202 336-12~\cite{etsi20233612}, provide a common semantic framework for representing power, energy, and environmental parameters, supporting interoperable telemetry, vendor-neutral monitoring, and scalable PEM deployment across heterogeneous RAN sites.

At the network level, PEM integration is enabled through cross-layer APIs that connect PEM with RAN control and management entities. The Service Management and Orchestration framework and the Non-Real-Time RAN Intelligent Controller serve as key integration points for policy-driven and AI-assisted optimization. Through these interfaces, PEM can ingest traffic forecasts, service priorities, mobility patterns, and operational intents such as planned carrier activation or peak-hour provisioning. In return, it exposes site-level power budgets, flexibility envelopes, battery constraints, and operating limits.

Near-real-time coordination can be supported through the Near-RT RIC and interfaces such as E2, enabling indirect energy-aware RAN actions including selective carrier activation, transmit-power adjustment, sleep-mode triggering, and traffic steering across cells. These mechanisms allow the RAN to exploit spatial and temporal variations in energy availability while maintaining QoS constraints and protecting mission-critical communication functions.

Different interfaces are required for different control timescales. High-frequency telemetry between PEM controllers and orchestration layers can use lightweight publish-subscribe protocols such as MQTT, while policy exchange and control signaling can use RESTful APIs or gRPC-based interfaces. Northbound, the RAN energy control plane interfaces with grid operators, aggregators, and market platforms. Protocols such as OpenADR and IEC 61850 profiles can support demand-response signals, flexibility requests, grid-service bids, and operational constraints, allowing aggregated RAN flexibility to be translated into standardized grid services.

A key architectural requirement is the enforcement of QoS-aware guardrails. Energy-control actions must remain subordinate to communication requirements, including minimum active carriers, coverage obligations, latency bounds, synchronization constraints, and backup-energy reserves. Feedback loops are therefore needed to reconcile energy optimization with network performance metrics and ensure that energy-aware adaptations remain transparent to end users.


AI-driven orchestration can further enhance PEM integration. Machine-learning models within the SMO, Non-RT RIC, or related management functions can forecast traffic, energy demand, renewable generation, and thermal behavior for proactive, constraint-aware packet scheduling. Overall, PEM support in 6G requires standardized telemetry, flexible APIs, RAN-aware orchestration, and interoperable grid-facing interfaces as the foundation for Telecoms VPP operation.

\section{Telecoms Virtual Power Plant}
A virtual power plant (VPP) digitally aggregates distributed energy resources, such as solar generation, battery storage, and flexible loads, to operate as a single controllable entity for balancing supply and demand and supporting grid stability. This paper introduces the concept of a \emph{Telecoms VPP}, which extends this principle to 6G infrastructure. In this vision, base stations and edge nodes equipped with renewables, storage, and intelligent power systems operate collectively as a grid-interactive energy resource. Underpinned by PEM, the telecoms VPP coordinates site-level consumption, storage, and generation to improve energy efficiency, reduce carbon intensity, and contribute flexibility to the grid. Its dual role is to optimize telecom operations without compromising connectivity and to deliver grid services such as demand response, renewable balancing, and ancillary support through aggregated flexibility. In doing so, it bridges the telecom and energy domains and transforms the RAN from a passive electricity consumer into an active participant in net-zero 6G operation.

\begin{figure}
\centering
\includegraphics[width=\columnwidth]{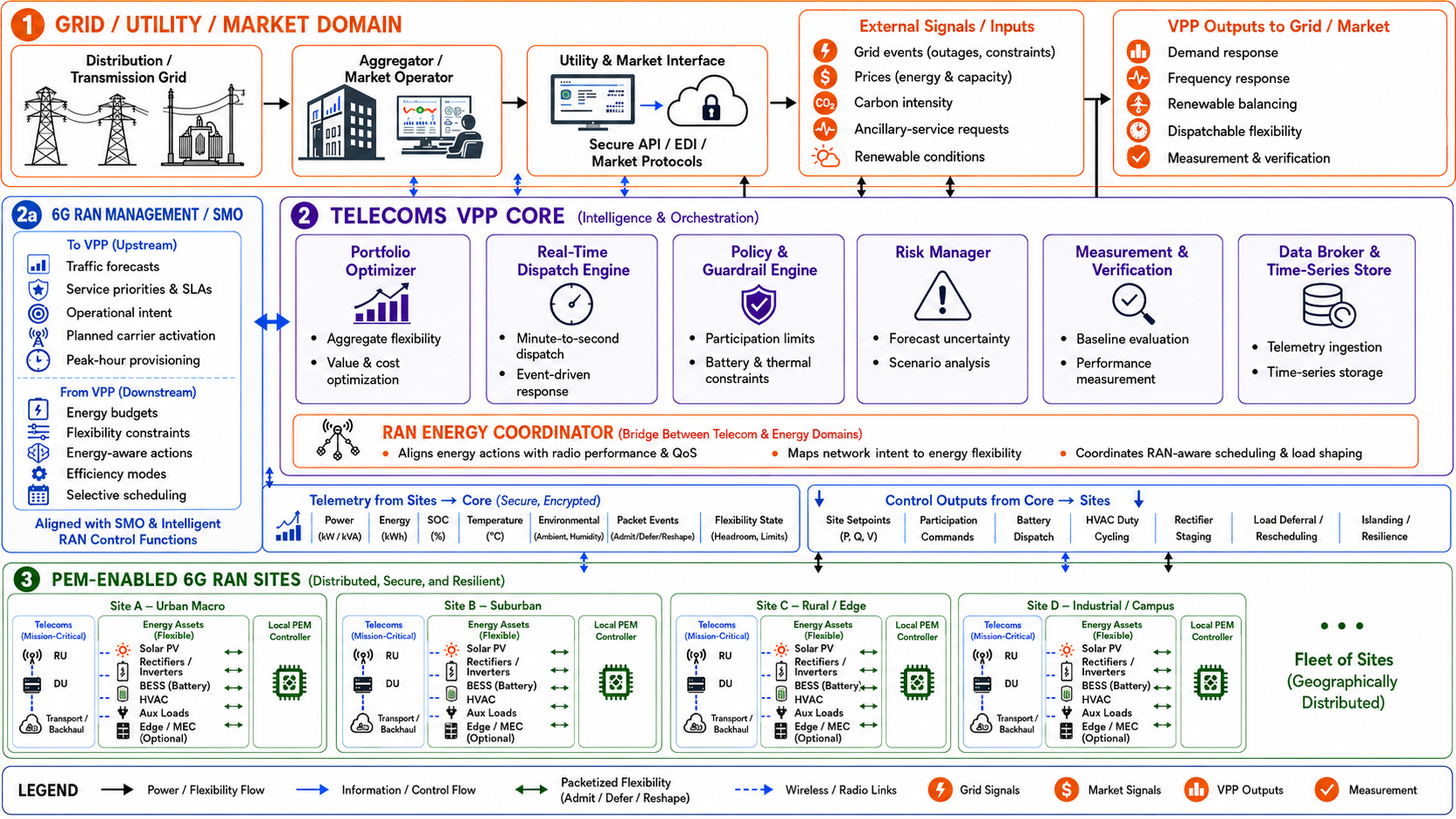}
\caption{Telecoms VPP architecture.}
\label{T_VPP}
\vspace{-1.5em}
\end{figure}

Building on this vision, Fig.~\ref{T_VPP} illustrates the proposed system-level architecture of a telecoms VPP, organized across the grid/market domain, the VPP core, RAN management functions, and PEM-enabled 6G RAN sites. At the site layer, each base station includes a local PEM controller connected to controllable electrical and thermal subsystems, including rectifiers, inverters, BESS, HVAC, thermal management, and on-site renewables where available. These assets provide the physical flexibility of the site and are coordinated through packetized actions that admit, defer, or reshape energy requests subject to safety constraints such as SOC limits, thermal bounds, and DC bus limits. Mission-critical communication functions, including RUs, DUs, and transport, remain protected by design and are not directly packetized. This local autonomy enables reliable operation during normal, degraded, or islanded conditions while generating high-resolution telemetry on power, energy, packet events, and environmental states.

Above the site layer, the Telecoms VPP core aggregates telemetry and flexibility states across geographically distributed RAN sites through secure interfaces. Its main functions include portfolio optimization, real-time dispatch, policy and guardrail enforcement, risk management, measurement and verification, and data management. The portfolio optimizer derives aggregate flexibility envelopes and schedules site participation over hourly to daily horizons, while the real-time dispatch engine converts grid events, market signals, or operator policies into site and cluster setpoints on shorter timescales. The policy and guardrail engine enforces telecom-specific constraints, including minimum connectivity, battery backup reserves, thermal protection, and per-site participation limits. The risk manager accounts for forecast uncertainty, operational risk, and resilience posture, while measurement and verification functions evaluate delivered flexibility against baselines for auditability, settlement, and reporting.

A distinguishing feature of the telecoms VPP is the RAN Energy Coordinator, which couples telecom service requirements with energy-control decisions. Through interfaces aligned with SMO and intelligent RAN control functions, the VPP ingests traffic forecasts, service priorities, SLAs, operational intents, planned carrier activation, and peak-hour provisioning requirements. These inputs are mapped into energy budgets, flexibility constraints, and site-level participation commands. Conversely, energy-side conditions such as limited renewable availability, reduced battery headroom, or grid constraints can inform energy-aware network actions, including efficiency modes, selective scheduling, and opportunistic scaling of non-critical processing. This feedback loop ensures that flexibility is primarily harvested from non-critical loads such as HVAC duty cycling, rectifier staging, battery dispatch, and deferrable workloads, while preserving deterministic behavior for real-time radio processing.

From the grid and market perspective, the telecoms VPP exposes aggregated RAN flexibility through a utility and market interface. It can support demand response by coordinating flexible-load deferral and controlled battery discharge to reduce net power draw. With sufficient storage headroom and appropriate interconnection capabilities, the same architecture can support frequency response and other ancillary services through fine-grained dispatch. These services are delivered through fleet-level aggregation rather than isolated site actions: the VPP selects participating clusters, distributes dispatch targets, compensates for site-level constraints, and verifies performance using high-resolution telemetry. This aggregation allows micro-flexibility at individual base stations to be translated into dispatchable capacity relevant to distribution and transmission system operators.

Overall, the telecoms VPP architecture converts PEM-enabled 6G RAN infrastructure into a coordinated flexibility resource that supports both internal operational optimization and external grid-service delivery. By combining local PEM autonomy, fleet-level orchestration, cross-layer RAN coordination, and grid-facing aggregation, it ensures that sustainability objectives are pursued without compromising connectivity, reliability, or safety. This architecture provides a practical pathway for integrating telecom networks into renewable-dominant power systems and positions the RAN as a strategic energy asset for net-zero 6G.

\section{Performance Evaluation}

A simulation framework is developed to evaluate PEM for a fleet of \(N=20\) 6G base-station sites over a 24-hour horizon with 5-minute resolution. Each site is modeled as a macro-cell-like RAN node with mission-critical communication loads, flexible auxiliary loads, thermal management, and, depending on the scenario, PV generation and battery energy storage. The same RAN traffic profiles and critical loads are used in all matched cases.
Six scenarios are evaluated in three matched pairs, each with and without PEM. The \emph{grid-only} case has no PV or storage, the \emph{conservative DER} case includes modest PV and battery capacity, and the \emph{net-zero-ready} case assumes larger PV and storage penetration. Critical RAN power includes idle radio, baseband, transport, monitoring, and traffic-dependent components, while flexible loads include HVAC/cooling, auxiliary loads, delay-tolerant edge workload, rectifier operation, and battery charge/discharge. Site heterogeneity is captured through variations in RAN power, traffic, HVAC limits, thermal parameters, PV capacity, and battery capacity.
PV generation follows a daytime solar profile with site-specific clouding. Grid carbon intensity and electricity price are modeled as time-varying exogenous signals, ranging from approximately \(0.12\) to \(0.55~\mathrm{kgCO_2/kWh}\) and \(0.05\) to \(0.25~\$/\mathrm{kWh}\), respectively. Thermal behavior is represented by a first-order indoor-temperature model, and battery operation follows SOC dynamics with efficiency, reserve, maximum SOC, and power-limit constraints.

For each infrastructure pair, the No-PEM case is first simulated to obtain a matched peak-demand reference. The PEM case then applies a peak-aware guardrail so that flexible-load recovery, battery charging, HVAC operation, and auxiliary-load scheduling do not increase peak grid demand. Under No-PEM operation, HVAC follows threshold-based control, PV is consumed locally, and batteries mainly absorb surplus PV. Under PEM, flexible loads issue implicit energy-packet requests that are admitted, deferred, or reshaped according to carbon intensity, price, renewable availability, thermal state, SOC, and peak constraints. Reported PEM results include terminal correction to ensure SOC-neutral and backlog-neutral comparison, preventing artificial gains from battery depletion or postponed computation. 

\begin{figure*}[!t]
\centering

\subfloat[]{%
\includegraphics[width=0.32\textwidth]{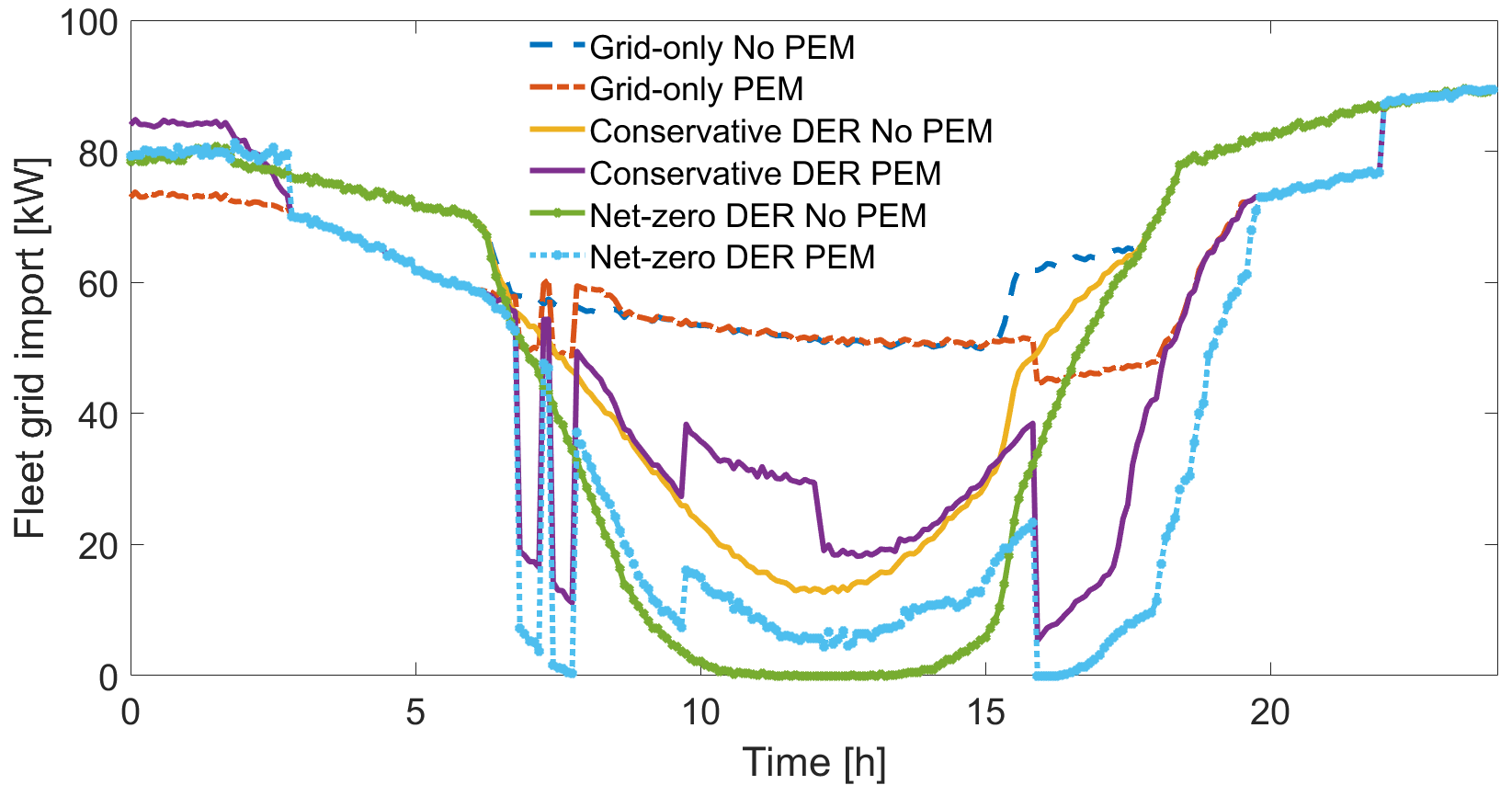}
\label{fleet_grid}}
\hfill
\subfloat[]{%
\includegraphics[width=0.32\textwidth]{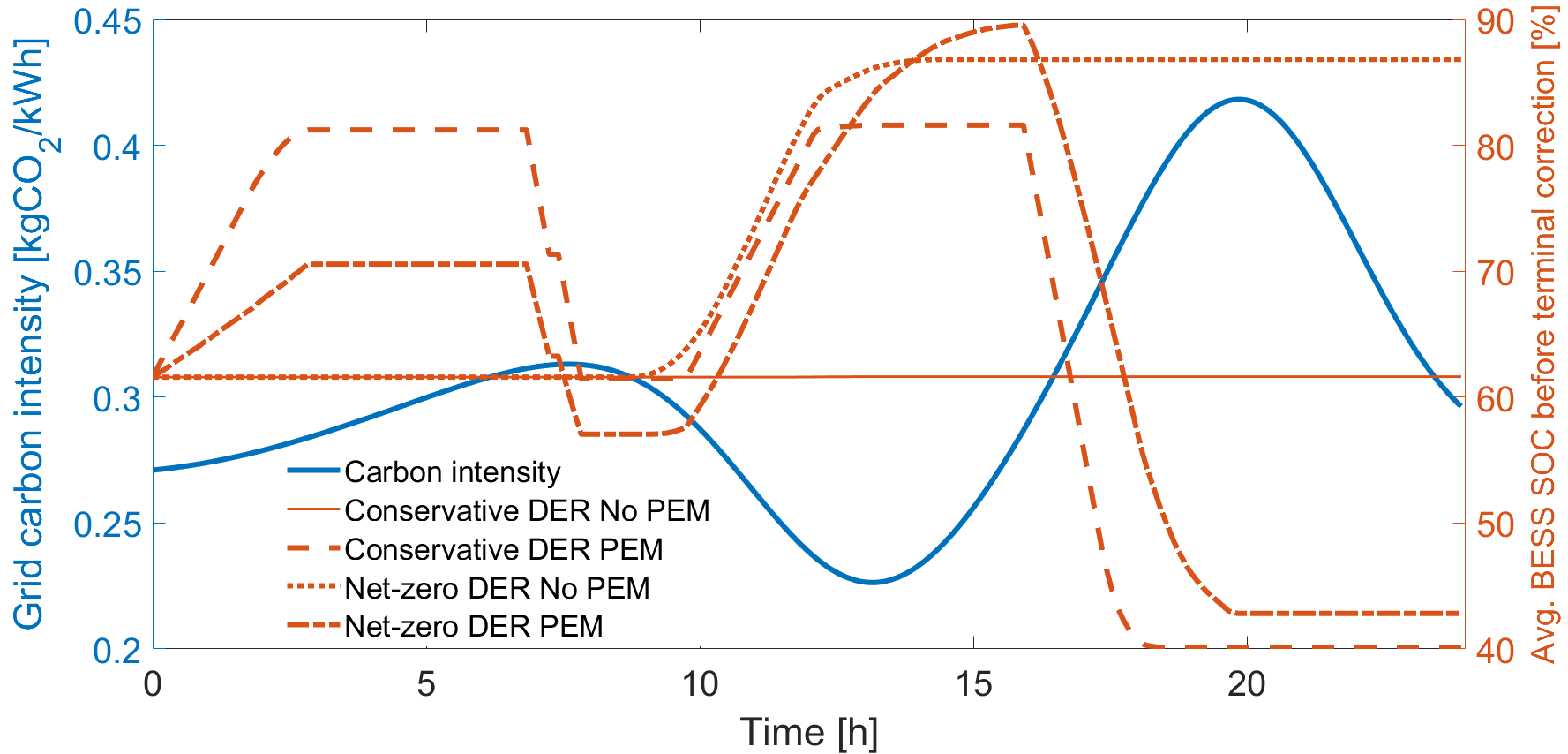}
\label{bess_soc}}
\hfill
\subfloat[]{%
\includegraphics[width=0.32\textwidth]{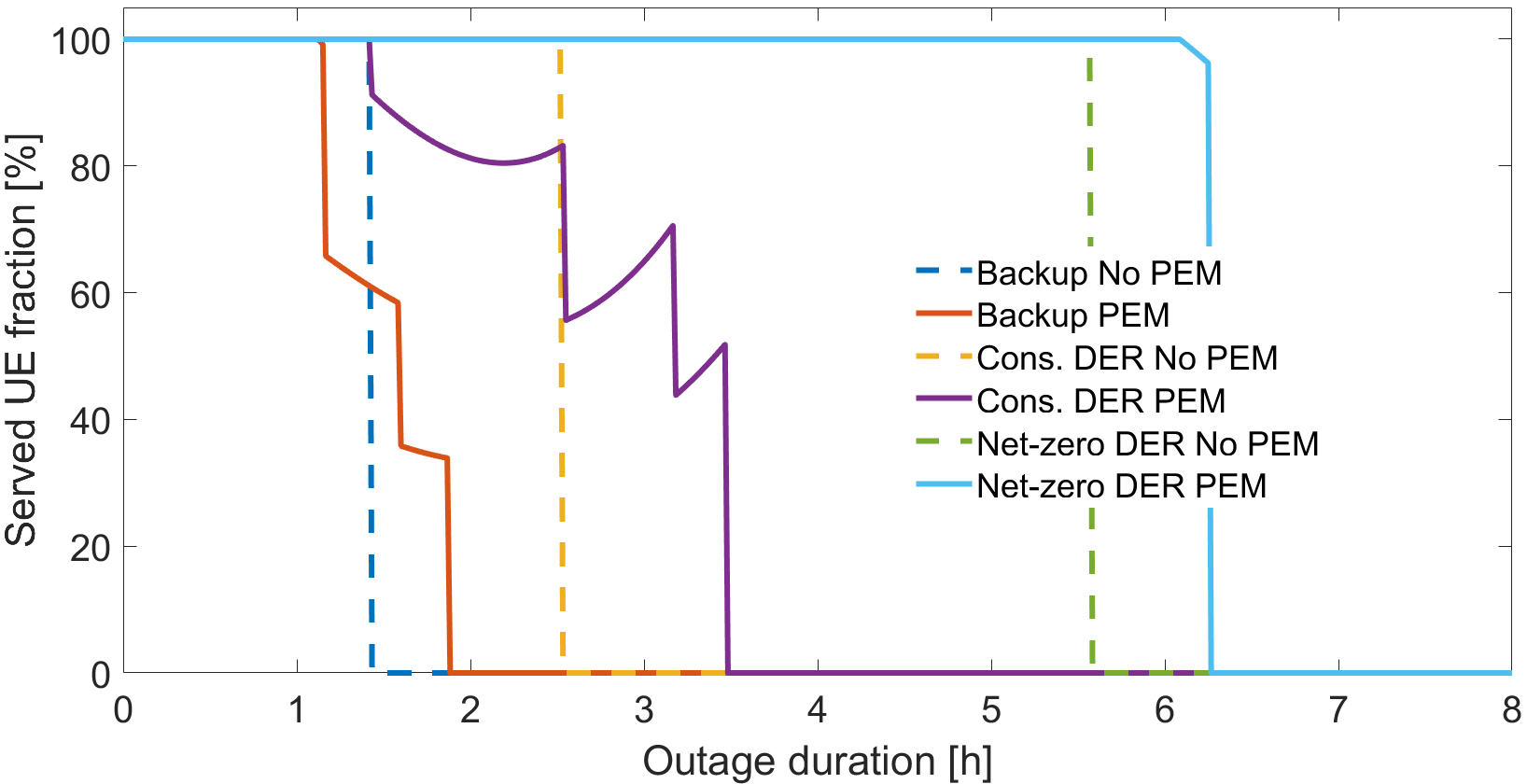}
\label{served_users}}
\vspace{-2mm}
\caption{Performance results under grid-only, conservative DER, and net-zero DER scenarios: (a) fleet-level grid import; (b) BESS state-of-charge trajectories under carbon-aware PEM operation; (c) served UEs during outage.}
\label{fig:pem_fleet_results}
\vspace{-4mm}
\end{figure*}

Fig.~\ref{fig:pem_fleet_results} and Table~\ref{tab:pem_six_scenario_results} jointly summarize the fleet-level impact of PEM. Fig.~\ref{fleet_grid} shows that DER integration reduces grid import during renewable-rich periods, while PEM further reshapes demand through flexible-load and storage coordination under the peak-aware guardrail. Fig.~\ref{bess_soc} illustrates carbon-aware storage behavior, where PEM preserves or charges batteries during favorable periods and discharges more during high-carbon intervals. The reported table values include terminal recharge correction, ensuring SOC-neutral and backlog-neutral comparison.
Table~\ref{tab:pem_six_scenario_results} confirms that PEM consistently reduces grid energy, carbon emissions, and operating cost across all matched scenarios. In the grid-only case, PEM reduces energy, carbon, and cost by \(7.83\%\), \(8.59\%\), and \(9.35\%\), respectively, showing that flexible-load coordination alone provides measurable benefit. In the conservative DER case, the corresponding reductions are \(7.30\%\), \(8.93\%\), and \(10.49\%\). Since PV self-consumption is already \(100\%\) with zero curtailment, these gains mainly arise from improved flexible-load and storage scheduling. The strongest results occur in the net-zero DER case, where PEM reduces energy by \(11.35\%\), carbon emissions by \(13.84\%\), and cost by \(16.45\%\). PEM also increases PV self-consumption from \(87.28\%\) to \(95.67\%\) and reduces PV curtailment from \(64.00\)~kWh to \(21.80\)~kWh. The peak remains slightly lower at \(89.46\)~kW compared with \(89.52\)~kW in the matched No-PEM cases, confirming that PEM improves energy, carbon, cost, and renewable utilization without increasing grid stress or relying on battery depletion.

A second simulation evaluates UE-level service continuity during an 8-hour energy outage with 1-minute resolution. Three matched infrastructure pairs are considered: backup-only, conservative DER, and net-zero-ready DER, each without and with PEM. The traffic profile is identical in each pair, with \(220\) UEs and a critical-service fraction of \(40\%\). In the No-PEM case, service collapses once available battery/PV supply can no longer support the load. In the PEM case, flexible loads are curtailed, delay-tolerant edge workload is deferred, and outage-mode admission control prioritizes critical UEs while preserving essential RAN functions.
Fig.~\ref{served_users} and Table~\ref{tab:ue_qos_outage_pem} show that PEM improves outage resilience by enabling graceful degradation rather than abrupt service collapse. In the backup-only case, PEM extends runtime from \(1.43\)~h to \(1.88\)~h, a \(31.4\%\) improvement, and increases critical UE-hours from \(90.58\) to \(123.13\), a \(35.93\%\) gain. With conservative DER, runtime increases from \(2.53\)~h to \(3.48\)~h, served UE-hours from \(455.49\) to \(524.39\), and critical UE-hours from \(182.20\) to \(256.14\). This corresponds to the largest runtime gain, \(37.5\%\), and the largest critical-service gain, \(40.58\%\), highlighting PEM's value under constrained energy availability. In the net-zero DER case, PEM extends runtime from \(5.58\)~h to \(6.27\)~h, increases average served UE fraction from \(69.65\%\) to \(78.13\%\), and reduces average outage probability from \(0.304\) to \(0.221\). Thus, PEM delivers the largest relative gains in energy-constrained sites and the strongest absolute service continuity in DER-rich deployments.

\begin{table*}[!t]
\centering
\caption{PEM-enabled 6G RAN fleet results with peak-aware and SOC-neutral operation.}
\label{tab:pem_six_scenario_results}
\scriptsize
\begin{tabular}{lcccccccc}
\hline
\textbf{Scenario} &
\textbf{Energy} &
\textbf{Carbon} &
\textbf{Cost} &
\textbf{Peak} &
\textbf{PV Self} &
\textbf{PV Curtailment} &
\textbf{SOC End} &
\textbf{Terminal Correction} \\
&
\textbf{[kWh]} &
\textbf{[kgCO$_2$]} &
\textbf{[\$]} &
\textbf{[kW]} &
\textbf{[\%]} &
\textbf{[kWh]} &
\textbf{[\%]} &
\textbf{[kWh]} \\
\hline
Grid-only No PEM        & 1630.28 & 509.70 & 203.16 & 89.52 & N/A    & 0.00  & N/A   & 0.00  \\
Grid-only PEM           & 1502.66 & 465.92 & 184.15 & 89.46 & N/A    & 0.00  & N/A   & 16.15 \\
Conservative DER No PEM & 1385.42 & 445.60 & 178.87 & 89.52 & 100.00 & 0.00  & 61.59 & 0.00  \\
Conservative DER PEM    & 1284.32 & 405.83 & 160.12 & 89.46 & 100.00 & 0.00  & 61.59 & 44.87 \\
Net-zero DER No PEM     & 1215.11 & 399.21 & 161.08 & 89.52 & 87.28  & 64.00 & 61.59 & 0.00  \\
Net-zero DER PEM        & 1077.25 & 343.96 & 134.59 & 89.46 & 95.67  & 21.80 & 61.59 & 59.13 \\
\hline
\end{tabular}
\vspace{-4mm}
\end{table*}

\begin{table*}[!t]
\centering
\caption{UE QoS performance during base station energy outage under priority-aware PEM}
\label{tab:ue_qos_outage_pem}
\scriptsize
\begin{tabular}{llccccccc}
\hline
\textbf{Infrastructure} &
\textbf{Policy} &
\textbf{Runtime} &
\textbf{Avg. Served UE} &
\textbf{Avg. Served Critical UE} &
\textbf{UE-hours} &
\textbf{Critical UE-hours} &
\textbf{Avg. Outage} &
\textbf{PV Used} \\
&
&
\textbf{[h]} &
\textbf{[\%]} &
\textbf{[\%]} &
\textbf{[hrs]} &
\textbf{[hrs]} &
\textbf{[probability]} &
\textbf{[kWh]} \\
\hline
Backup-only      & No PEM & 1.43 & 17.88 & 17.88 & 226.46 &  90.58 & 0.821 &  0.00 \\
Backup-only      & PEM    & 1.88 & 19.13 & 23.03 & 247.64 & 123.13 & 0.816 &  0.00 \\
Conservative DER & No PEM & 2.53 & 31.60 & 31.60 & 455.49 & 182.20 & 0.684 &  3.83 \\
Conservative DER & PEM    & 3.48 & 36.16 & 43.45 & 524.39 & 256.14 & 0.659 &  4.69 \\
Net-zero DER     & No PEM & 5.58 & 69.65 & 69.65 & 885.34 & 354.13 & 0.304 & 12.91 \\
Net-zero DER     & PEM    & 6.27 & 78.13 & 78.17 & 956.56 & 382.77 & 0.221 & 12.73 \\
\hline
\end{tabular}
\vspace{-4mm}
\end{table*}

\section{Key Challenges}

\begin{itemize}

\item \textit{Integration with RAN orchestration} -- PEM must be integrated with Open RAN SMO, Non-RT RIC, and Near-RT RIC functions without disrupting existing RAN control loops. Energy-aware actions such as traffic steering, sleep modes, carrier activation, and edge-workload placement must remain aligned with performance requirements.


\item \textit{Renewable and storage sizing} -- PEM benefits depend on the sizing of PV, battery storage, and power-electronic interfaces. Oversized assets increase cost and underutilization, while undersized assets limit carbon reduction and resilience. Joint planning of DER capacity and PEM control is therefore needed for cost-effective deployment.


\item \textit{AI-driven scheduling} -- PEM depends on forecasts of traffic, renewables, price, carbon intensity, battery state, and thermal conditions. Explainable AI-based scheduling is needed to manage uncertainty while ensuring safe and predictable telecom operation.

\item \textit{Standardization} -- PEM-enabled RAN requires interoperable telemetry, control APIs, and energy-packet signaling across telecom equipment, power systems, batteries, cooling units, and grid-facing platforms. Standardization is needed for site-level PEM control, orchestration interfaces, and grid-service interactions.

\item \textit{Market participation} -- Aggregating PEM-enabled base stations into a telecoms VPP requires market models that account for coverage obligations, SLA protection, outage risk, and battery reserves. Cellular-specific flexibility models, bidding strategies, and settlement mechanisms are needed for reliable grid-service participation.

\item \textit{Extension beyond the RAN} -- PEM can also extend to edge data centers, aggregation sites, core-network facilities, transport nodes, and flexible UE or IoT energy tasks. A broader framework is needed to coordinate energy flexibility across the full  infrastructure stack.

\end{itemize}

\section{Concluding Remarks}

Net-zero 6G requires energy management beyond incremental RAN efficiency gains. As AI-native operation, edge intelligence, and renewable-assisted deployments increase network energy complexity, this paper introduces PEM as a framework for transforming 6G base stations from passive electricity consumers into energy-aware, grid-interactive assets.

The proposed PEM-enabled RAN architecture separates mission-critical communication functions from flexible energy loads, enabling cooling, rectifiers, batteries, auxiliary loads, and delay-tolerant workloads to be coordinated without compromising connectivity. The Telecoms VPP concept further aggregates PEM-enabled sites to optimize energy use, carbon intensity, cost, resilience, and grid-service capability. In the net-zero DER case, PEM reduces fleet energy use by \(11.35\%\), carbon emissions by \(13.84\%\), and operating cost by \(16.45\%\), while maintaining peak-demand discipline. It also increases PV self-consumption from \(87.28\%\) to \(95.67\%\) and reduces curtailment by \(65.93\%\).
The outage study confirms PEM's resilience value. In the conservative DER case, PEM extends runtime by \(37.5\%\) and increases critical UE-hours by \(40.58\%\); in the backup-only case, it improves runtime by \(31.4\%\) and critical UE-hours by \(35.93\%\). These results show that PEM can align sustainability, resilience, and grid flexibility in 6G. Future work should address field validation, standardized PEM interfaces, telecom-specific flexibility markets, and AI-assisted control policies that preserve communication QoS while delivering verifiable energy flexibility.

\bibliographystyle{IEEEtran}
\bibliography{6G_Copy.bib}

\end{document}